\documentclass[preprint,showpacs,preprintnumbers,amsmath,amssymb,floats]{revtex4}
\usepackage{graphicx}
\usepackage{dcolumn}
\usepackage{bm}

\newcommand{\fk}{f_{\mathbf{k}}}

\newcommand{\be}{\begin{eqnarray}}
\newcommand{\ee}{\end{eqnarray}}

\newcommand{\bd}{b^{\dagger}}
\newcommand{\cd}{c^{\dagger}}
\newcommand{\dd}{d^{\dagger}}

\begin{document}

\title{Collective Modes in the Loop Ordered Phase of Cuprates}

\author{Yan He and C.M. Varma}

\affiliation{Department of Physics, University of California, Riverside, CA}
\date{\today}
\begin{abstract}
We show that the two branches of collective modes discovered recently in under-doped Cuprates with huge spectral
weight are a necessary consequence of the loop-current state. Such a state has been shown in earlier experiments to be consistent with the symmetry of the order parameter competing with superconductivity in four families of Cuprates. We also predict a third  branch of excitations and suggest techniques to discover it. Using parameters to fit the observed modes, we show that the direction of the effective moments in the ground state lies in a cone at an angle to the c-axis as observed in experiments.
 \end{abstract}
\maketitle

It is generally agreed that the most important theoretical problem posed by the revolutionary discoveries in the Cuprates is an understanding of the normal states above $T_c(x)$, which present new paradigms in the organization of matter.  Specifically these states are the pseudogap state at low dopings and the non-fermi-liquid state at intermediate dopings. A novel broken symmetry with a large order parameter, (estimated \cite{li} to be $0.3\mu_B/unit cell$ in Hg(1201) with $T_c = 61 K$), has been discovered in four different families of cuprates \cite{fauque, mook, li, baledent, kaminski} below the pseudogap temperature $T^*(x)$, with $T^*(x) \to 0$ at $x \to x_c$, the quantum critical point. This state can therefore be regarded as a universal feature of the phase diagram of the Cuprates. The symmetry of the state with one important discrepancy is consistent with the spatial and time-reversal symmetry of a state predicted to arise by formation of ordered loop currents in a microscopic model for the Cuprates \cite{cmv1, cmv2}. The marginal non- fermi-liquid properties have been derived from the quantum-critical fluctuations of this state\cite{aji-cmv} and the coupling of the fermions to such fluctuations has been shown \cite{aji-shekhter-cmv} to promote d-wave superconductivity with the right scale of $T_c$ and the superconducting gap. It is therefore necessary for a general acceptance of the loop ordered state as the Rosetta stone whose deciphering leads to an understanding of all the other universal properties of Cuprates, that every aspect of it be convincingly understood. Two crucial aspects towards this goal are achieved in this paper.

If there is an unusual order parameter in the pseudogap state, there should also be unusual collective modes there. Recently, inelastic neutron scattering has discovered \cite{li2, lithesis} two branches of  weakly dispersive collective modes  throughout the Brillouin zone in several underdoped $Hg1201$ for $ T \lesssim T^*(x)$. The integrated strength of these modes at low $T$ is estimated to be over 20 times larger than the intensity in the so-called $(\pi,\pi)$ resonance \cite{li2, lithesis}. Nearly dispersion free modes have already been indirectly inferred through their dominant coupling to the fermions in ARPES \cite{lanzara} and optical conductivity experiments \cite{hemmen}. The energy and the spectral strength of the newly discovered modes make them the most likely candidates to explain the ARPES and optical conductivity experiments.

In this paper we show that three branches of excitations must be present in the loop-current ordered state but one of them can not be discovered by neutron scattering. We calculate the frequency dependence of these modes through introducing the quantum-mechancial version of the Ashkin-Teller model which describes the statistical mechanics of the loop ordered state. We predict that the third branch can be discovered through Raman or inelastic x-ray scattering. Moreover, we resolve the major discrepancy of the order parameter found in polarized neutron Bragg scattering from the predictions of the classical model. We show that the quantum-model gives moments lying on a cone around the c-axis and calculate the angle of the cone from the parameters needed to fit the observed collective modes. This angle is in agreement with the experiments. We also make several predictions for future experiments.

\begin{figure}[b]
\centerline{\includegraphics[width=0.8\textwidth]{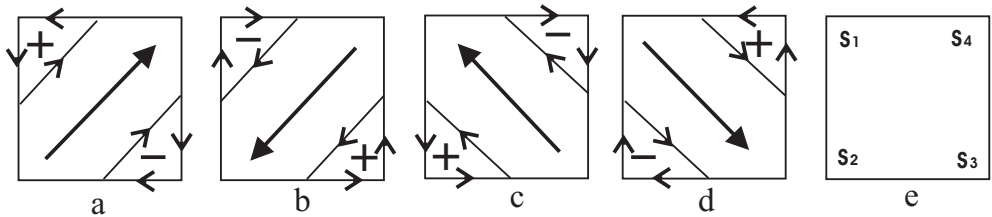}}
\caption{The four Possible "classical" domains of the loop ordered state are shown. Fig. (e) presents the notation discussed in the text for the location of the moments within the unit-cell. }
\label{configs}
\end{figure}

The four possible "classical" domains of the loop current ordered state are shown in Fig. (\ref{configs}). The "classical" order parameter \cite{shekhter} may be characterized by the {\it anapole} vector \cite{zeldovich} ${\bf L}$
\be
\label{def L}
{\bf L} = \int_{cell} d^2r  ({\bf M}({\bf r})\times {\bf \hat{r}}) \approx  \sum_{\mu}{\bf M_{\mu}} \times {{\bf r}}_{\mu}
\ee
where the moment distribution ${\bf M}({\bf r})$ is formed due to the currents on the four O-Cu-O triangles per unit-cell as shown. An approximate representation of ${\bf L}$ is given by ${\bf M}({\bf r})\approx {\bf M_{\mu}} \delta^2({\bf r} -{\bf r}_{\mu})$, where ${\bf r}_{\mu}$, ${\mu} = 1,..,4$ are the location of the four "sites" in any cell at the centroid of the moment distribution. These sites are labelled $S1,..,S4$ in Fig.(\ref{configs}e). The orbital moments ${\bf M_{\mu}}$ are either up or down or zero. The four {\it classical} domains of Fig.(\ref{configs}e) may be represented by the four values of the angle $\theta = \pi/4, 3\pi/4, 5\pi/4, 7\pi/4$ that ${\bf L}$ makes with the $\hat{x}-$axis.


The classical statistical mechanics of the Loop-Current state may be derived from the
Ashkin-Teller model, which is given in terms of a pair of Ising spin per unit-cell $\sigma_{i3}, \tau_{i3}$, whose eigenvalues, $\pm 1$ specify the $x$ and $y$ components of the direction of the vector ${\bf L}$ and whose eigenvectors are $|10,10\rangle,\, |01,10\rangle,\, |01,01\rangle,\, |10,01\rangle$; the first two numbers specify eigenstates of $\sigma_{i3}$ and the second two of $\tau_{i3}$.
\be
\label{classAT}
H_{AT} =  \sum_{(ij)}\frac{J}{4} ( \sigma_{i3}\sigma_{j3} + \tau_{i3}\tau_{j3}) + \frac{J_4}{4}  \sigma_{i3}\tau_{i3} \sigma_{j3}\tau_{j3}.
\ee

Quantum fluctuations among the four possible directions of order together with dissipation lead \cite{aji-cmv} to a scale invariant spectrum which leads to the observed Marginal Fermi-liquid properties in the quantum-critical regime. These quantum fluctuations are specified in the basis of the eigenstates of the operator ${\bf L}$.
The 4 classical loop current states are eigenvalues of operators ${\bf L}_i=(L_{i,x},L_{i,y})$ defined at site $i$. According to Eq. (8) of Ref.(\onlinecite{aji-shekhter-cmv}),
\be
\label{Lx,y}
(L_x+iL_y) |{\theta}>=\exp^{i\theta} |{\theta}>
\ee
The quantum fluctuations are generated from a term in the microscopic hamiltonian:
\be
H_Q = \sum_i 2t {\bf L}_{zi}^2
\ee
where ${\bf L}_{zi}$ is the generator of rotations in the space of the four states of ${\bf L}_i$, i.e
\be
U(\pm \frac{\pi}{2}) |\Theta> = \exp(\pm i\frac{\pi}{2}{\bf L}_{zi})|\theta> = |\theta \pm \pi/2>
\ee
It can then be shown that
\be
\label{Lz}
L_z &=& (1/2)\big[-(1+i)U + U^2 + (-1+i)U^3\big] \\
L_z^2 &=& 5-2\sqrt 2(U+U^{\dagger}),
\ee
From the operations of $U+U^{\dagger}$ on the basis set chosen in Eq.(\ref{classAT}), it follows that
\be
U+U^{\dagger} = \frac{1}{\sqrt 2}\left(\begin{array}{cccc} 0 & 1+i & 0 & 1-i \\ 1-i & 0 & 1+i & 0 \\ 0 & 1-i & 0 & 1+i \\ 1+i & 0 & 1-i & 0
\end{array}\right)
\ee
Note that $(U + U^{\dagger})$ rotates the states $|\theta>$ both clockwise and anti-clockwise by $\pi/2$ but with matrix elements which are complex. This is represented in terms of the AT model by an operator $(\sigma_1 +\sigma_2+ \tau_1 +\tau_2)$.  It is worth noting that in the present problem the kinetic energy does not break any lattice symmetry since it is derived from a term $L_{zi}^2$ in the original model. This is  unlike the case of the transverse field Ising model.

A more convenient representation for calculations on the problem is $SU(4)$; we introduce the direct products,
\be
S^i=\sigma^i\otimes I,\quad T^i=I\otimes\tau^i,\quad K^{ij}=\sigma^i\otimes\tau^j.
\ee
The quantum AT model in this representation is
\be
\label{fullH}
H_{QAT}= \sum_{i} t (S^1_i+T^1_i +S^2_i+T^2_i) + t' (K_i^{11} + K_i^{22})
-\sum_{\langle i,j\rangle} \frac{J}{4} (S^3_i S^3_j + T^3_i T^3_j) + \frac{J_4}{4} K^{33}_i K^{33}_j.
\ee
The second term above expresses the effect of $U^2$ which rotates ${\bf L}$ by $\pi$. The magnitude of $4t/J$ (in the present notation) was estimated \cite{aji-shekhter-cmv} to be of $O(1)$. It is hard to estimate $t'$, which we leave as a free parameter.

We can parameterize the most general quantum state by the three pairs of angles $\theta_i, \phi_i, i=1,2,3$:
\be
\label{wavefn}
|\psi\rangle &=&\cos\frac{\theta_1}{2}\Big(\cos\frac{\theta_2}{2}|10, 10\rangle
+\sin\frac{\theta_2}{2}e^{i\phi_2}|01, 10\rangle\Big) \\ \nonumber
&+&\sin\frac{\theta_1}{2} e^{i\phi_1}\Big(\cos\frac{\theta_3}{2} |10, 01\rangle
+\sin\frac{\theta_3}{2} e^{i\phi_3}|01, 01\rangle\Big)
\ee
{\it Special Case $J_4=0$}:
To understand the results for the general case, it is useful first to consider the simple case, $J_4 =0$, when we simply have two decoupled transverse field Ising models. The linear collective modes  for the transverse field Ising model were calculated by de Gennes \cite{deGennes}.
In this case we have $\phi_i=0$ for $i=1,2,3$ and $\theta_2=\theta_3$. Minimizing the Hamiltonian we find $\sin\theta_1=\sin\theta_2=\sin\theta=-t/J$. The ground
state is
\be
\label{directproduct gs}
|G\rangle&=&\Big(\cos\frac{\theta}{2}|10\rangle
+\sin\frac{\theta}{2}|01\rangle\Big)_{\sigma} \bigotimes \Big(
\cos\frac{\theta}{2}|10\rangle
+\sin\frac{\theta}{2}|01\rangle \Big)_{\tau}
\ee
There are three eigenvalues at any $k$, two of which are degenerate,
\be
\label{eigenvalueJ4=0}
\omega_{1,2} &=& 2(J^2 - t^2 \fk)^{1/2};~\fk = (\cos(k_xa) + \cos(k_ya))/2, \\ \nonumber
\omega_{3} &=& 4J
\ee
$\omega_{1,2}$ correspond to the propagation of the flipped states in the $\sigma$ and $\tau$ sectors respectively due to the (quantum) transverse-fields. $\omega_3$ is the energy for flipping the angle of ${\bf L}$ by $\pi$. The modes in this approximation were derived earlier \cite{cmv1} by a simple procedure and called the "mixing modes" and the "phase mode" in a semi-classical calculation with no four-spin coupling.\\
{\it General case}:
Since the 1 and the 2 components of all operators appear symmetrically, $\phi_i = \pi/4$ for all $i$ in the ground state. This leaves 3 real coefficients in the general wavefunction. Also we are interested only in the case $J_1=J_2$ and the Hamiltonian is invariant under interchange of $\sigma$ and $\tau$. So the general wave-function, as we have verified by explicit calculation, has only two other parameters. It turns out through minimizing numerically the ground state energy that the wave-function (\ref{directproduct gs}) with just one parameter (beside $\phi$) is a good approximate (correct to within a few percent) ground state wave-function for the interesting range of parameters.

To obtain the collective modes in the general case, it is convenient to introduce the Holstein-Primakoff transformation generalized to $SU(4)$. We introduce three boson operators $b$,$c$, and $d$ such that
\be
&&S^1=U c+\cd U+\bd d+\dd b,\quad\quad S^3=1-2\cd c-2\dd d\\
&&T^1=U b+\bd U+\cd d+\dd c,\quad\quad T^3=1-2\bd b-2\dd d\\
&&K^{11}=U d+\dd U+\bd c+\cd b,\quad\quad K^{33}=1-2\bd b-2\cd c\\
&&K^{13}=U c+\cd U-\bd d-\dd b,\quad\quad K^{31}=U b+\bd U-\cd d-\dd c
\ee
with $U=(1-\bd b-\cd c-\dd d)^{1/2}$. Here $\bd,\,\cd,\,\dd$ are the creation operators of states $|01, 10\rangle$, $|10, 01\rangle$ and $|01,01\rangle$ respectively.

We transform the Hamiltonian to the Boson representation, perform a generalized Bogoliubov transformation and calculate the ground state wavefunction and the eigenvalues and the eigenvectors of the excitations. The general results can only be obtained through numerical diagonalization of a matrix. We compare such results below with the experiments. But first we give the perturbative corrections for $J_4/J\ll1, t' =0$ to the eigenvalues given in Eq. (\ref{eigenvalueJ4=0}) for the case $J_4 = 0$:
\be
&&\delta(\omega_1^2)^{(1)}=4J^2\left[2-(3+\fk)\frac{t^2}{J^2}
+(1+3\fk/2)\frac{t^4}{J^4}-\fk\frac{t^6}{2J^6}\right]\,\frac{J_4}{J}\nonumber\\
&&\delta(\omega_2^2)^{(1)}=4J^2\left[2-(1-\fk)\frac{t^2}{J^2}
-(1+3\fk/2)\frac{t^4}{J^4}+\fk\frac{t^6}{2J^6}\right]\,\frac{J_4}{J}\nonumber\\
&&\delta(\omega_3^2)^{(1)}=8J^2\left[4\frac{t^2}{J^2}-(4+\fk)\frac{t^4}{J^4}\right]\,\frac{J_4}{J}
\ee
Note that the degeneracy of the modes $\omega_1, \omega_2$ is lifted by $J_4 \ne 0$. This reflects that operator $K_{33}$ couples the $\sigma$ and $\tau$ modes. For $t' =0$, the two modes represent the two linear combinations $\sigma \pm \tau$. For $t' \ne 0$, these modes themselves couple and a more complicated linear combination varying with ${\bf k}$ results, as we see through our general results obtained numerically. We can also determine the correction perturbatively in $t'$ for $J_4 =0$ to get
\be
&&\delta(\omega_1^2)^{(1)}=4J^2\left[1-(2-3\fk)\frac{t^2}{2J^2}
+\fk\frac{t^4}{2J^4}\right]\,\frac{t'}{J}\nonumber\\
&&\delta(\omega_2^2)^{(1)}=4J^2\left[-1+(2+5\fk)\frac{t^2}{2J^2}
-\fk\frac{t^4}{2J^4}\right]\,\frac{t'}{J}\nonumber\\
&&\delta(\omega_3^2)^{(1)}=32t^2\frac{t'}{J}\nonumber
\ee

Now we present the results for the eigenvalues of the collective modes calculated numerically for any given $J, J_4, t, t'$.  For both $J_4, J >0$, the dispersion of the upper mode is smaller than the lower. For $J >0, J_4 <0, t' =0$, the energy of the upper mode at $k =0$ progressively decreases as $k$ increases and it must (from the phase diagram of the classical AT model) \cite{baxter} go to $0$ at the zone boundary. With $t' \ne 0$, there is a level repulsion and anti-crossing of the two modes.

\begin{figure}
\centerline{\includegraphics[width=0.6\textwidth]{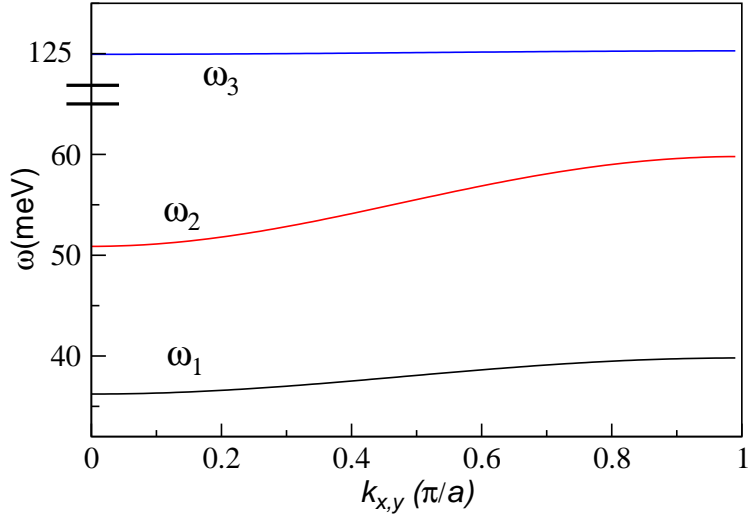}}
\caption{The collective mode dispersions of Ashkin-Teller model with negative as a function of $k_x$ along the $k_x=k_y$ direction. The black curve stands for $\omega_1$ and red curve stands for $\omega_2$. The parameters are taken to fit the experimental results in $Hg1201, T_c = 65 K$.. The blue curve gives the calculated third mode.}
\label{dispersion}
\end{figure}

Choosing parameters to reproduce the experiments in $Hg1201$ with $T_c = 65 K$, we show the results in Fig. (\ref{dispersion}). In the experiments, one of the collective modes is at $40\pm 5$ meV another one at $50 \pm 5$ meV at $k=0$. The dispersion across the Brillouin zone in both the $(11)$ and the $(10)$ directions for the higher energy mode is $5 \pm 5$ meV, while the lower mode is even less dispersive. We also have the very important constraint from the thermodynamics of the AT model \cite{baxter, gronsleth} that to have a transition with no divergence in the specific heat  $-1 < J_4/J < 0, J>0$. In Fig. (\ref{dispersion}), the following parameters are used:
\be
t=5.5\,\mbox{meV},\, J=33.2\,\mbox{meV},\,J_4=-0.3J,\,t'=-1.7t,\, \sin\theta\approx-0.37
\ee
We calculate  $\sin\theta\approx-0.37$ for these parameters. In this case the dispersion width of the first mode is about $10\%$ of the energy gap and second one is around $5\%$ of the energy gap. The result is shown in figure \ref{dispersion}. These dispersions change with doping and we use a doping where a lot of data is available. We can estimate the exact transition temperature $T_c$ from using Baxter's \cite{baxter} result, $\exp(-J_4/2T_c) = \sinh(J/2T_c)$ to be about 200 Kelvin compared to the experimental value of about 250 Kelvin.

The high energy mode whose eigenvalue is $\omega_3$ cannot be observed by polarized neutron scattering since, it involves change of angular momentum by 2 but should be observable by other tools.

{\bf Semi-Classical Angle of the Effective Moments}\\
The effective direction of the moments is deduced in polarized neutron diffraction experiments by measuring the intensity of the spin-flipped scattered neutrons at a
 a Bragg vector ${\bf Q}$ by sending in the neutron beam at various different polarization directions  and using standard selection rules \cite{neutrons}.
 By this method the direction of ${\bf M}$ has been determined to be $45 \pm 20$ degrees to the normal to the plane both in a sample of $Y(123)$ and $Hg(1201)$.

The scattering of neutrons from ordered orbital moments leads to a few considerations not usually encountered in scattering due to ordered spins. The moments at the four "sites" in Fig. (\ref{configs}e) are orbital moments and so are represented by the transformation properties of $S=1$ spins, rather than $S=1/2$ spins. The neutron spins are represented by the usual spin-1/2 Pauli matrices ${\bf \sigma}_n$, which change the angular momentum of the system by $\pm 1$ in the spin-flip process. So at any "site" the spin-fip due to scattering of a neutron makes the transitions $(1 \to 0), (-1 \to 0)$ and {\it vice-versa} but there is no transition from the $(1 \to -1)$ or vice-versa. This should be contrasted with ordered spin-systems, for example for ${\bf S}=1/2$, where the neutrons scatter from states of $S_z=-1/2$ to $S_z=1/2$ and vice-versa.

To calculate the neutron scattering intensity, we must find the expectation value of $\langle{\bf M}({\bf Q})\rangle$ in the ground state. ${\bf M}({\bf Q})$ is the Fourier transform of the real space magnetic moment operator. The representation of ${\bf M}({\bf r})$ in terms of the order parameter ${\bf L}$ is obtained by inverting Eq. (\ref{def L}) to get
\be
\label{M-qm}
{\bf M}_i({\bf r}_{\mu})={\bf L}_i\times {{\bf r}}_{\mu}/{{\bf r}}_{\mu}^2
\ee
Transforming to momentum space
\be
\langle{\bf M}({\bf Q})\rangle = \sum_i e^{i {\bf Q. R}_i}\sum_{\mu} \frac{\langle {\bf L}_i\rangle \times {\bf r}_{\mu}}{{\bf r}_{\mu}^2} e^{i {\bf Q. r}_{\mu}},
\ee
where the expectation value is taken in the ground state wave-function, Eq.(\ref{directproduct gs}), which we have found to be accurate to a few percent. Using Eq. (\ref{Lz}) for the matrix representation of ${\bf L}$,  we find
\be
\langle L_x \rangle = \langle L_y \rangle = \frac{1}{\sqrt{2}}r_0^{-1} \cos \theta \\ \nonumber
\langle L_z \rangle = \frac{1}{\sqrt 2} r_0^{-1} \sin \theta(\frac{\sin \theta}{2\sqrt 2} -1).
\ee
$r_0$ is the magnitude of ${\bf r}_{\mu}$ measured from the center of the cell.
Using these we find
\be
\langle M_x({\bf Q})\rangle &=& - \frac{1}{\sqrt{2}} \sin \theta(\frac{\sin \theta}{2\sqrt 2} -1) \Big(e^{i {\bf Q. r}_1} -e^{i {\bf Q. r}_{2}}-e^{i {\bf Q. r}_{3}}+e^{i {\bf Q. r}_{4}}\Big), \\ \nonumber
\langle M_y({\bf Q})\rangle &=& - \frac{1}{\sqrt{2}} \sin \theta(\frac{\sin \theta}{2\sqrt 2} -1)\Big(e^{i {\bf Q. r}_1} +e^{i {\bf Q. r}_{2}}-e^{i {\bf Q. r}_{3}}-e^{i {\bf Q. r}_{4}}\Big), \\ \nonumber
\langle M_z({\bf Q})\rangle &=& \sqrt{2} \cos \theta \Big(e^{i {\bf Q. r}_1} -e^{i {\bf Q. r}_{3}}\Big), \nonumber
\ee
Here we have taken the magnitude of ${\bf M}$ to be unity since we wish to compare with experiments only the ratio of ${\bf M}$ in different directions.

For the momentum transfer ${\bf Q} = (0, a^*, c^*)$, as in the experiment and for the domain chosen (a in Fig.(\ref{configs})
\be
\langle{\bf M}\rangle = 2 i\sin (a^* r_y) \Big(\frac{1}{\sqrt{2}}\sin \theta (\frac{\sin \theta}{\sqrt 2} -2), 0, \sqrt{2} \cos \theta\Big)
\ee
In the other domains $M_x, M_y$ switch. The tilt angle to the normal is
\be
\phi = \arctan\Big[\frac{\sin \theta (\frac{\sin \theta}{\sqrt 2} -2)}{2\cos \theta}\Big]
\ee
For fitting the spectrum of collective modes as shown in Fig.(\ref{dispersion}), we needed $\sin \theta \approx -0.37$. This yields $\phi \approx 25$ degrees. If we increase $t$ by $20\%$, the upper mode disperses by about 10 meV, which is also in the experimental range. Then we can get $\phi \approx 33$ degreea. The experimental range \cite{fauque, lithesis} is estimated to be $40 \pm 20$degrees. This means that spin-orbit scattering\cite{spins} or additions\cite{weber} to the simplest two-dimensional model may not be required to understand the dependence of the intensity of the neutron scattering on the polarization direction.

A prediction, beside that of the third mode which may be discovered through Raman or inelastic x-ray scattering, is that  the frequency of all three branches of the collective modes should $\to 0$ for $q \to 0$ as $T \to T^*(x)$, because the transition is second order. As already noted \cite{li}, the observed ARPES spectra in underdoped cuprates

To summarize, it is important for an acceptable theory of the Cuprates that the universal features of the phase diagram and the principal excitations in each part of it be understood consistently and quantitatively from a common set of idea.
This paper provides a crucial contribution towards this goal through showing that the dominant new fluctuations in the pseudogap phase are necessary excitations of the loop-ordered phase which has symmetry consistent with the uiversal order parameter observed in underdoped cuprates. An additional crucial result is that we find the direction of the effective ordered moments to be consistent with experiments through the same quantum-model from which the observed collective modes are derived. We also provide several new testable predictions.

{\it Acknowledgements}: We wish to thank Philippe Bourges, Martin Greven and Yuan Li for discussion of experimental results prior to publication and to Vivek Aji, Philippe Bourges, Vladimir Cvetcovic, Martin Greven and Thierry Giamarchi (TG) for many discussions and important corrections on the preliminary manuscript. In particular TG pointed out an error in an earlier derivation of $L_z$.  This work is supported partially by NSF grant DMR-0906530.

\end{document}